\documentclass[aps,prl,reprint,superscriptaddress]{revtex4-2}
\usepackage{amsfonts, amssymb, amsmath,graphicx}
\usepackage{multirow}
\usepackage{dcolumn}
\usepackage{bm}
\usepackage{color}
\usepackage[colorlinks=true,allcolors=blue]{hyperref}

\begin{document}

\title{Interaction-induced non-Hermitian topological phases from a dynamical gauge field}

\author{W. N. Faugno}
\affiliation{Advanced Institute for Materials Research (WPI-AIMR), Tohoku University, Sendai 980-8577, Japan}
\author{Tomoki Ozawa}
\affiliation{Advanced Institute for Materials Research (WPI-AIMR), Tohoku University, Sendai 980-8577, Japan}

\date{\today}

\begin{abstract}
We present a minimal non-Hermitian model where a topologically nontrivial complex energy spectrum is induced by inter-particle interactions. Our model consists of a one-dimensional chain with a dynamical non-Hermitian gauge field with density dependence. The model is topologically trivial for a single particle system, but exhibits nontrivial non-Hermitian topology with a point gap when two or more particles are present in the system. We construct an effective doublon model to describe the nontrivial topology in the presence of two particles, which quantitatively agrees with the full interacting model. Our model can be realized by modulating hoppings of the Hatano-Nelson model; we provide a concrete Floquet protocol to realize the model in atomic and optical settings.
\end{abstract}

\maketitle

Non-Hermitian Hamiltonians have been found to host a rich variety of topological phases~\cite{Kawabata19,Esaki11,Schomerus13,Weimann17,Bahari17,St-Jean17,Lieu18a,Parto18,Harari18,Yin18,Lieu18b,Shen18,Philip18,Chen18}. While some non-Hermitian phases are direct analogues of Hermitian phases, there are many uniquely non-Hermitian phases. These result when the system has a point gap in the complex energy plane, which allows for non-trivial winding of the energy spectrum. Non-Hermitian topology manifests in an analogous bulk-boundary correspondence known as the skin effect wherein a macroscopic number of states localize at the boundary~\cite{Lee16,Alvarez18,Kunst18,Kawabata18,Tao19}. These phenomena have been primarily investigated as single particle effects.

Comparatively little work has been done to understand the role of correlations and interactions in non-Hermitian topological phases~\cite{Yoshida19,Nakagawa20,Kawabata22,McClarty19,Mook21}. Interactions have played an important role in Hermitian topological physics, giving rise to many paradigmatic phases including the fractional quantum Hall effect and quantum spin liquids. Such strongly interacting systems have led to many advancements in physics, including developments in gauge theories. Given the richness of Hermitian interacting systems, it remains to be seen how analogous non-Hermitian interactions can enrich the topology of open systems. Experimentally, two body loss terms are ubiquitous in optical lattices and photonics with an increasing degree of control, further motivating investigations of the topology of many body open systems.

In this letter, we report on a minimal 1D non-Hermitian model exhibiting interaction induced topology. We demonstrate that our model is topologically trivial for a single particle, but gains a non-trivial winding number in the complex energy plane for two or more particles. We characterize the spectrum by the clustering properties of the eigenstates. This leads us to derive an effective SSH model of Doublons with an emergent sublattice symmetry, which quantitatively captures the complex energy ring of the full spectrum. The winding number of the interacting model corresponds with the winding of this effective model analogous to interaction induced topology in Hermitian systems~\cite{Stepanenko20,Olekhno20,Salerno2020}. We conclude by proposing a two-frequency Floquet protocol that realizes our model as an effective Hamiltonian. As an intermediate step, this Floquet protocol realizes a Hatano-Nelson model.

\begin{figure}[htbp]
     \centering
    \includegraphics[width=\columnwidth]{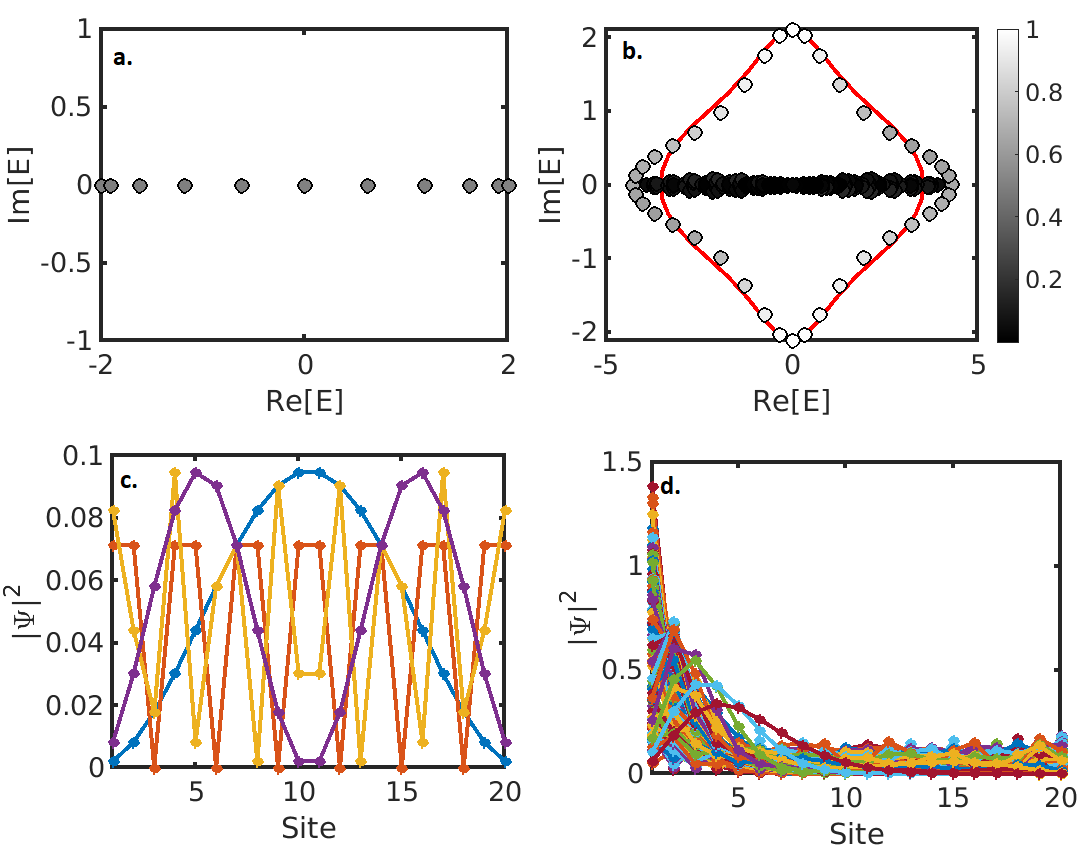}
    \caption{Summary of key results. Panels a and b are energy spectra for periodic boundary conditions plotted in the complex plane for 1 and 2 particles, respectively. For one particle, the spectrum is real while for two particles, the spectrum is complex with a point gap. The coloring in panel b is obtained projecting each eigenstate onto the subspace of basis states where particles lie on the same site or adjacent sites. The red line is the spectrum obtained from the effective doublon model described below. Panel c and d are the real space profiles of the eigenstates in the open boundary geometry for 1 and 2 particles respectively. For one particle, eigenstates are typical standing waves while for two particles we observe a skin effect. The chosen parameters are $t=1$, $\gamma_L=1.5$ and $\gamma_R=0$ for a lattice with 20 sites.}
    \label{fig:spectra}
     
\end{figure}

\textit{Model.---}
Our model consists of bosons populating a 1D chain with hoppings dependent on the gradient of the density, which can be interpreted as a density-dependent synthetic dynamical gauge field. The Hamiltonian of our system is
\begin{align}
    H = \sum_j &a^\dagger_{j+1}\bigg[-t + i\gamma_R(n_{j+1}-n_j)\bigg]a_j \notag \\
    &+a^\dagger_j\bigg[-t + i\gamma_L(n_j-n_{j+1})\bigg]a_{j+1} 
    \label{eqtn:H}
\end{align}
where $a_j$ and $a^\dagger_j$ are bosonic annihilation and creation operators, respectively, $t$ is the single particle hopping parameter, $n_j$ is the density operator $a^\dagger_ja_j$ on the $j$th site, and $\gamma_{R/L}$ are the couplings to the gauge field for right and left hoppings. To realize a non-Hermitian model, we take $\gamma_R \neq \gamma_L^*$ in similar fashion to the Hatano-Nelson model~\cite{Hatano96,Hatano98}. Description in terms of non-Hermitian Hamiltonians can be obtained through a post-selection procedure on quantum trajectories \cite{Nakagawa20,Schafer2020}. Later we present a concrete experimental protocol combining quantum trajectory and Floquet theory to realize this Hamiltonian. In the present model, when there is only one particle present in the system, the density terms are identically zero. Therefore, for a single particle the Hamiltonian is Hermitian and corresponds to a free boson. The single particle spectrum is shown in Fig. \ref{fig:spectra}a which reproduces the free boson result for a periodic chain of length $L=20$.

\begin{figure}[htbp]
     \centering
     \includegraphics[width=.5\textwidth]{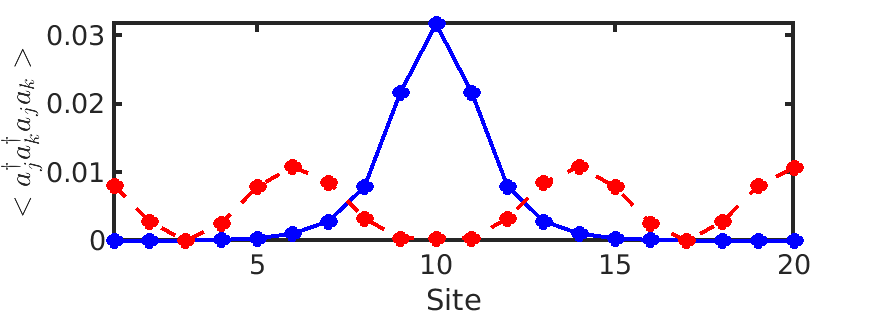}
    \caption{Four point correlator $\langle a_j^\dagger a_k^\dagger a_ja_k\rangle$ for $k=10$ with periodic boundary conditions. The solid line is representative of states with corresponding energies on the ring while the dashed line is representative of states with nearly real eigenenergies.} \label{fig:correlator}
     
\end{figure}

\textit{Exact Diagonalization.---}
Let us contrast this result with the two particle spectrum shown in Fig \ref{fig:spectra}b. We consider fixed particle number as the Hamiltonian has U(1) symmetry. Physically, particle number is conserved between quantum jumps during which the non-Hermitian Hamiltonian description is valid. Here we find that the energy spectrum consists of a sector where energies are nearly real and a sector consisting of a ring of complex energies. We project each eigenstate into the subset of basis states where particles lie on the same site or adjacent sites, and find that the states with complex energies largely lie in this subspace while those with nearly real energies have almost zero weight in this subspace. To further characterize this separation we calculate the correlator $a_j^\dagger a_k^\dagger a_ja_k$ for each eigenstate. Representative correlators for complex energy states and nearly real energy states are shown as a function of $j$ for a fixed $k$ in Fig \ref{fig:correlator}. This correlator confirms that the states with complex energy occur when the particles cluster while those with nearly real energies occur when the particles separate.

\begin{figure}
    \centering
    \includegraphics[width=.5\textwidth]{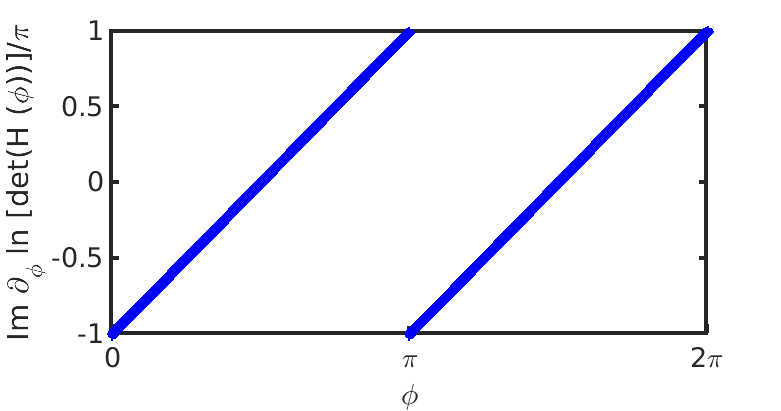}
    \caption{Plot demonstrating the nontrivial winding number. A jump from 1 to -1 increases the winding by 1 while a jump from -1 to 1 descreases the winding by 1.}
    \label{fig:winding}
\end{figure}

The energy spectrum has a point gap indicating the presence of nontrivial topology. To verify the nontrivial topology, we calculate the winding number following the flux insertion procedure outlined in Ref \cite{Zong18}. We define
$H(\phi)$ by multiplying $e^{-i\phi}$ ($e^{i\phi}$) to the boundary hopping term in the first (second) term of Eq.(\ref{eqtn:H}),
where $\phi$ is the strength of the inserted magnetic flux. We then calculate
\begin{equation}
    \frac{1}{\pi}\Im\bigg[\partial_\phi\ln{\det[{H(\phi) - \delta I}]}\bigg]
    \label{eq:winding}
\end{equation}
as a function $\phi$, where $\Im [\cdot]$ stands for the imaginary part, and $I$ is the identity matrix. The signed number of jumps in this quantity give the winding of the phase about the point $\delta$, chosen to lie within the point gap, in the complex plane. In Fig \ref{fig:winding}, we plot this quantity versus the flux, $\phi$, and clearly see that it jumps twice as the flux is tuned from 0 to $2\pi$, giving a winding number of 2 and confirming that the system is topologically non-trivial when there are two particles. 

\begin{figure}
    \centering
    \includegraphics[width = .3\textwidth]{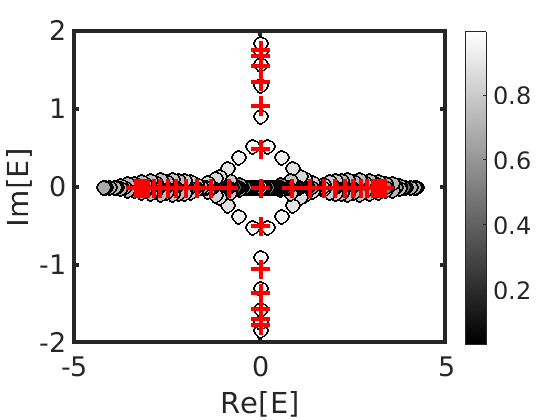}
    \caption{Energy spectrum for open boundary geometry. The parameters are the same as in Fig.~\ref{fig:spectra} and the coloring of points is obtained through the same projection method. Red crosses are obtained from the effective Doublon SSH model.}
    \label{fig:OBCspectrum}
\end{figure}

A non-trivial winding number implies the existence of the skin effect in the open boundary geometry. The open-boundary eigenstates are plotted in Figs \ref{fig:spectra}c and \ref{fig:spectra}d for the one and two particle systems respectively. The one particle eigenstates are exactly those obtained for a free Boson model while the two particle eigenstates demonstrate a clear skin effect. In fact, all states will localize on the edge for strong enough gauge coupling. For completeness, we present the energy spectrum in the open boundary geometry in Fig \ref{fig:OBCspectrum}, which shows that the spectrum does not cleanly separate along particle clustering properties as both particles localize on the edge. Unlike the single-particle Hatano-Nelson model, the spectrum in the open boundary condition does not lie on the real axis only; we discuss below the origin of this complex spectrum.

\begin{figure}
    \centering
    \includegraphics[width=0.5\textwidth]{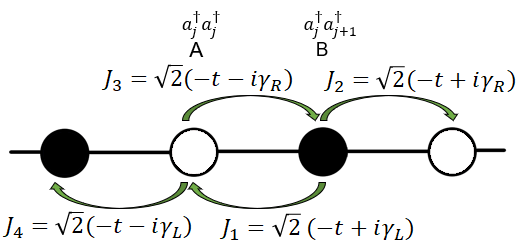}
    \caption{Diagram of the effective doublon SSH model. Sublattice $A$ maps to basis states with particles on the same site while sublattice $B$ maps to basis states with particles on adjacent sites.}
    \label{fig:SSHdia}
\end{figure}

\textit{Effective Doublon Model.---}
To understand the topology of the system, we derive an effective doublon model that captures the physics of the complex energy ring. We obtain this by restricting our basis to states where the particles lie on the same site or on adjacent sites. 
The effective doublon model consists of two sublattices; we define sublattice A to be the set of states with particles on the same site, and sublattice B to be the set of states with two particles occupying adjacent sites, as shown in the diagram in Fig \ref{fig:SSHdia}.
We define the creation operator of a particle in $j$-th site of sublattice A as $a_{A,j}^\dagger \equiv a^\dagger_j a^\dagger_j$ and that of sublattice B as $a_{B,j}^\dagger \equiv a^\dagger_j a^\dagger_{j+1}$.
The resulting effective model is given by the Hamiltonian
\begin{align}
    H_{Doublon} &= \sum_j \Psi^\dagger_{j,j}
    \begin{pmatrix}
        0 & J_1 \\
        J_3 & 0 
    \end{pmatrix}\Psi_{j,j}
+\Psi^\dagger_{j+1,j}
\begin{pmatrix}
    0 & J_2 \\
    J_4 & 0 
\end{pmatrix}\Psi_{j+1,j},
\label{eq:doublonH}
\end{align}
where $\Psi_{j,l}^\dagger \equiv \left( a_{A,j}^\dagger, a_{B,l}^\dagger\right)$ and
$J_1=\sqrt{2}(-t+i\gamma_L)$, $J_2=\sqrt{2}(-t+i\gamma_R)$, $J_3=\sqrt{2}(-t-i\gamma_R)$, and $J_4=\sqrt{2}(-t-i\gamma_L)$,
which is an SSH model with asymmetric hoppings on a chain of length $2L$.
We note that the effective doublon model has emergent sublattice symmetry in which hopping within the same sublattice is absent. The energy spectrum of this model is plotted in red on Figs.~\ref{fig:spectra}b and \ref{fig:OBCspectrum} for periodic and open boundaries, respectively, demonstrating that key features of the many body spectrum are captured by this single particle doublon model. The accuracy of the Doublon model increases for stronger density-dependent hopping.

To characterize the topology of this effective theory, we calculate the winding number of the energy spectrum in momentum space.
The Fourier transform of the effective Hamiltonian is
\begin{equation}
    \sum_k \tilde{\Psi}^\dagger_{k,k}
    \begin{pmatrix}
        0 & J_1+J_2e^{-ik} \\
        J_3+J_4e^{ik} & 0 
    \end{pmatrix}\tilde{\Psi}_{k,k}
\end{equation}
with
$\tilde{\Psi}_{k,k} \equiv (\tilde{a}_{A,k},\tilde{a}_{B,k})$, where $\tilde{a}_{A,k}$ and $\tilde{a}_{B,k}$ are the Fourier transforms of $a_{A,j}$ and $a_{B,j}$.
The details for determining the topology of such a system with sublattice symmetry are outlined in Ref \cite{Kawabata19} where we have the upper diagonal matrix $H^+=J_1+J_2e^{-ik}$ and the lower diagonal matrix $H^-=J_3+J_4e^{ik}$. If the point gap in the many body system is well-defined, the winding of the full system without considering the sublattice symmetry is twice  the winding of the doublon Hamiltonian. The doublon model tells us approximately which points in the complex energy plane we should consider to find non-trivial winding in the full many-body system.
Additionally, the point gaps in the many body system are often larger than in the doublon effective theory, providing additional points about which there is non-trivial winding.

In the Hatano-Nelson model, the spectrum under open boundary conditions is real implying there exists an imaginary gauge transformation between the Hatano-Nelson Hamiltonian and a Hermitian Hamiltonian~\cite{Hatano96,Hatano98,Okuma20b}. In general the energy spectrum of $H_{Doublon}$ has both imaginary and real energies, implying that the Hamiltonian cannot be made Hermitian by an imaginary gauge transformation, i.e. our effective doublon model is not related to any Hermitian matrix by a similarity transformation.
With a similarity transformation similar to the one used for the ordinary Hatano-Nelson model~\cite{Okuma20b}, one can transform our effective doublon Hamiltonian under an open boundary condition to a tri-diagonal form:
\begin{align}
    H_{Doublon} \sim
    \begin{pmatrix}
    0 & \sqrt{J_1 J_3} & 0 & \cdots \\
    \sqrt{J_1 J_3} & 0 & \sqrt{J_2 J_4} & \cdots \\
    0 & \sqrt{J_2 J_4} & 0 &\cdots \\
    \vdots & \vdots & \vdots & \ddots
    \end{pmatrix}.
\end{align}
This matrix is real and symmetric when both $\sqrt{J_1 J_3}$ and $\sqrt{J_2 J_4}$ are real, which gives a sufficient condition for the reality of the spectrum.
When $\gamma_L$ and $\gamma_R$ are real, which is the situation relevant in this paper, $J_2 J_4 = (J_1 J_3)^*$, in which case we numerically confirm that $J_1 J_3 > 0$ gives the necessary and sufficient condition for the spectrum being entirely real.

\textit{Floquet Protocols.---}
To realize our model, we have identified a Floquet protocol whose effective Hamiltonian can be identified with Eq. \ref{eqtn:H}. This Floquet protocol is inspired by previous efforts towards realizing Hermitian density dependent gauge fields in the setting of optical lattices ~\cite{Keilmann11,Rapp12,Greschner14,Greschner15,Meinert16,Wang20,Weitenberg21}. The system consists of a static Bose-Hubbard Hamiltonian and a time periodic Hatano-Nelson model.

We first describe how to realize Hatano-Nelson model from a Floquet protocol, which consists of a repeated three step process as presented in Ref. \cite{Goldman14} with total frequency $\Omega=2\pi/T_\Omega$.\cite{Zhang21} The time-dependent Hamiltonian is given as
\begin{equation}
    H(t) = \Delta \sum_j (a^\dagger_{j+1}a_j + a_{j+1}a^\dagger_j) + U\sum_j n_j(n_j-1)+ V(t)
    \label{eq:timedepH}
\end{equation}
where $\Delta$ is the hopping parameter, $U$ is the strength of the interaction and $V(t)$ is the modulation given by
\begin{equation}
    V(t) = 
    \begin{cases}
      \Delta_1 \sum_j a^\dagger_{j+1}a_j + a_{j+1}a^\dagger_j, & 0 \le t < T_\Omega/3 \\
      \sum_j i\mu_ja^\dagger_ja_j, & T_\Omega/3 \le t < 2T_\Omega/3 \\
      0, & 2T_\Omega/3 \le t < T_\Omega
    \end{cases}\label{eqtn:3stepHN}
\end{equation}
corresponding to a free gas of bosons with periodically modulated hopping and site dependent loss $\mu_j$. Note $t$ is defined mod $T_\Omega$. We obtain an effective Hamiltonian to order $1/\Omega$ as
\begin{equation}
\begin{split}
    H_{eff} = \sum_j \big[\Delta-\frac{\pi\Delta_1}{27\Omega}(\mu_j-\mu_{j+1})\big]a^\dagger_{j+1}a_j\\
    + \big[\Delta-\frac{\pi\Delta_1}{27\Omega}(\mu_{j+1}-\mu_{j})\big]a^\dagger_{j}a_{j+1} + U\sum_j n_j(n_j-1)
    \end{split}
\end{equation}
Choosing the site-dependent loss strength to be $\mu_j = j\mu_0$, we obtain a Hatano-Nelson model. A non-Hermitian Hamiltonian with site dependent loss can be implemented in ultracold gases by applying near-resonant light with position-dependent intensity~\cite{Takasu20}. Modulating this electric field in time as proposed above should give rise to the effective time-dependent Hamiltonian in Eqs. \ref{eq:timedepH} and \ref{eqtn:3stepHN}.

By further modulating the hopping of the effective Hatano-Nelson Hamiltonian, we obtain the desired hopping as we now show. The full time dependent Hamiltonian is given by
\begin{equation}
\begin{split}
    H(t) =& \Delta\sum_j \bigg[a^\dagger_{j+1}a_j + a^\dagger_ja_{j+1}\bigg] + U\sum_ja^\dagger_ja_j(a^\dagger_ja_j-1)\\
    + \sin&(\omega t)\sum_j\bigg[\Delta_Ra^\dagger_{j+1}a_j + \Delta_La^\dagger_ja_{j+1} \bigg]
\end{split}
\end{equation}
where $\Delta$ is the hopping parameter, $U$ is the interaction strength, $\omega$ is the frequency of the drive, and $\Delta_R \neq \Delta_L^*$ describes a non-Hermitian drive analogous to a Hatano-Nelson model. We assume $\omega \ll \Omega$ so that in the time scale of $1/\omega$ the system is effectively described by a static Hatano-Nelson model. From the Magnus expansion\cite{Maricq82,Grozdanov88}, we can obtain the effective Hamiltonian to order $1/\omega$
\begin{align}
    H_{\text{eff}} &= \sum_j a^\dagger_{j+1}\bigg[\Delta-\frac{2}{i\hbar\omega} U\Delta_R(n_j-n_{j+1})\bigg]a_j \notag \\
    &+ a^\dagger_{j}\bigg[\Delta-\frac{2}{i\hbar\omega} U\Delta_L(n_{j+1}-n_{j})\bigg]a_{j+1} \notag \\
    &+ U\sum_j a_j^\dagger a_j (a_j^\dagger a_j - 1)
\end{align}
which maps to our original system with $\Delta = -t$, $\gamma_L = \frac{2}{\hbar\omega}U\Delta_L$ and $\gamma_R = \frac{2}{\hbar\omega}U\Delta_R$. We note here if one considers this Floquet protocol for fermions instead of bosons, the Hatano-Nelson model can be realized from the fast frequency oscillation, but the density-dependent hopping is absent as particles cannot lie on the same site. This Floquet protocol introduces an additional interaction term in the effective Hamiltonian. The physics of our proposed model remains unchanged for sufficiently small interaction strength $U$.

Besides one dimensional quantum systems with controlled modulation, such as ultracold atomic gases, the above protocols can also be realized by mapping the models to two dimensional classical systems, where density-induced Hermitian models have been previously investigated~\cite{Stepanenko20,Olekhno20}.

\textit{Conclusion.---}
We have presented a minimal model where the interaction induces non-trivial non-Hermitian topology. The density-dependent non-Hermitian gauge field played a crucial role in the emergence of two-body topological phases. Our work paves a way toward understanding the role of gauge fields in non-Hermitian many-body systems. In Hermitian systems, effects of gauge fields are pronounced in two or higher dimensions, giving rise to exotic many-body phases such as fractional quantum Hall phases and topological order. It is therefore of great interest to extend the study of many-body physics in non-Hermitian systems under gauge fields to two and higher dimensions, in which we expect interplay of topological orders and non-Hermiticity.
The Floquet protocol we provide also elucidated that such non-Hermitian many-body physics can be studied in experimentally viable setups under suitable time modulations.
Furthermore, density-dependent gauge fields are simplest examples of dynamical gauge fields where gauge fields do not take externally fixed values. In Hermitian systems, dynamical gauge fields play a fundamental role in understanding a wide variety of systems, from high-energy to condensed matter physics. Their relations to non-Hermitian physics has been little explored; our work opens an avenue toward exploring this uncharted field. 

\text{\it{Acknowledgements}} - The authors would like to thank Zeyu Zhang and Mikael C Rechtsman for discussions leading to the Floquet protocol proposed in the paper.
This work is supported by JSPS KAKENHI Grant No. JP20H01845, JST PRESTO Grant No. JPMJPR19L2, and JST CREST Grant No.JPMJCR19T1.

\bibliography{biblio.bib}

\begin{thebibliography}{44}%
\makeatletter
\providecommand \@ifxundefined [1]{%
 \@ifx{#1\undefined}
}%
\providecommand \@ifnum [1]{%
 \ifnum #1\expandafter \@firstoftwo
 \else \expandafter \@secondoftwo
 \fi
}%
\providecommand \@ifx [1]{%
 \ifx #1\expandafter \@firstoftwo
 \else \expandafter \@secondoftwo
 \fi
}%
\providecommand \natexlab [1]{#1}%
\providecommand \enquote  [1]{``#1''}%
\providecommand \bibnamefont  [1]{#1}%
\providecommand \bibfnamefont [1]{#1}%
\providecommand \citenamefont [1]{#1}%
\providecommand \href@noop [0]{\@secondoftwo}%
\providecommand \href [0]{\begingroup \@sanitize@url \@href}%
\providecommand \@href[1]{\@@startlink{#1}\@@href}%
\providecommand \@@href[1]{\endgroup#1\@@endlink}%
\providecommand \@sanitize@url [0]{\catcode `\\12\catcode `\$12\catcode
  `\&12\catcode `\#12\catcode `\^12\catcode `\_12\catcode `\%12\relax}%
\providecommand \@@startlink[1]{}%
\providecommand \@@endlink[0]{}%
\providecommand \url  [0]{\begingroup\@sanitize@url \@url }%
\providecommand \@url [1]{\endgroup\@href {#1}{\urlprefix }}%
\providecommand \urlprefix  [0]{URL }%
\providecommand \Eprint [0]{\href }%
\providecommand \doibase [0]{https://doi.org/}%
\providecommand \selectlanguage [0]{\@gobble}%
\providecommand \bibinfo  [0]{\@secondoftwo}%
\providecommand \bibfield  [0]{\@secondoftwo}%
\providecommand \translation [1]{[#1]}%
\providecommand \BibitemOpen [0]{}%
\providecommand \bibitemStop [0]{}%
\providecommand \bibitemNoStop [0]{.\EOS\space}%
\providecommand \EOS [0]{\spacefactor3000\relax}%
\providecommand \BibitemShut  [1]{\csname bibitem#1\endcsname}%
\let\auto@bib@innerbib\@empty
\bibitem [{\citenamefont {Kawabata}\ \emph {et~al.}(2019)\citenamefont
  {Kawabata}, \citenamefont {Shiozaki}, \citenamefont {Ueda},\ and\
  \citenamefont {Sato}}]{Kawabata19}%
  \BibitemOpen
  \bibfield  {author} {\bibinfo {author} {\bibfnamefont {K.}~\bibnamefont
  {Kawabata}}, \bibinfo {author} {\bibfnamefont {K.}~\bibnamefont {Shiozaki}},
  \bibinfo {author} {\bibfnamefont {M.}~\bibnamefont {Ueda}},\ and\ \bibinfo
  {author} {\bibfnamefont {M.}~\bibnamefont {Sato}},\ }\bibfield  {title}
  {\bibinfo {title} {Symmetry and topology in non-hermitian physics},\ }\href
  {https://doi.org/10.1103/PhysRevX.9.041015} {\bibfield  {journal} {\bibinfo
  {journal} {Phys. Rev. X}\ }\textbf {\bibinfo {volume} {9}},\ \bibinfo {pages}
  {041015} (\bibinfo {year} {2019})}\BibitemShut {NoStop}%
\bibitem [{\citenamefont {Esaki}\ \emph {et~al.}(2011)\citenamefont {Esaki},
  \citenamefont {Sato}, \citenamefont {Hasebe},\ and\ \citenamefont
  {Kohmoto}}]{Esaki11}%
  \BibitemOpen
  \bibfield  {author} {\bibinfo {author} {\bibfnamefont {K.}~\bibnamefont
  {Esaki}}, \bibinfo {author} {\bibfnamefont {M.}~\bibnamefont {Sato}},
  \bibinfo {author} {\bibfnamefont {K.}~\bibnamefont {Hasebe}},\ and\ \bibinfo
  {author} {\bibfnamefont {M.}~\bibnamefont {Kohmoto}},\ }\bibfield  {title}
  {\bibinfo {title} {Edge states and topological phases in non-hermitian
  systems},\ }\href {https://doi.org/10.1103/PhysRevB.84.205128} {\bibfield
  {journal} {\bibinfo  {journal} {Phys. Rev. B}\ }\textbf {\bibinfo {volume}
  {84}},\ \bibinfo {pages} {205128} (\bibinfo {year} {2011})}\BibitemShut
  {NoStop}%
\bibitem [{\citenamefont {Schomerus}(2013)}]{Schomerus13}%
  \BibitemOpen
  \bibfield  {author} {\bibinfo {author} {\bibfnamefont {H.}~\bibnamefont
  {Schomerus}},\ }\bibfield  {title} {\bibinfo {title} {Topologically protected
  midgap states in complex photonic lattices},\ }\href
  {https://doi.org/10.1364/OL.38.001912} {\bibfield  {journal} {\bibinfo
  {journal} {Opt. Lett.}\ }\textbf {\bibinfo {volume} {38}},\ \bibinfo {pages}
  {1912} (\bibinfo {year} {2013})}\BibitemShut {NoStop}%
\bibitem [{\citenamefont {AU~Weimann}\ \emph {et~al.}(2017)\citenamefont
  {AU~Weimann}, \citenamefont {Kremer}, \citenamefont {Plotnik}, \citenamefont
  {Lumer}, \citenamefont {Nolte}, \citenamefont {Makris}, \citenamefont
  {Segev}, \citenamefont {Rechtsman},\ and\ \citenamefont
  {Szameit}}]{Weimann17}%
  \BibitemOpen
  \bibfield  {author} {\bibinfo {author} {\bibfnamefont {S.}~\bibnamefont
  {AU~Weimann}}, \bibinfo {author} {\bibfnamefont {M.}~\bibnamefont {Kremer}},
  \bibinfo {author} {\bibfnamefont {Y.}~\bibnamefont {Plotnik}}, \bibinfo
  {author} {\bibfnamefont {Y.}~\bibnamefont {Lumer}}, \bibinfo {author}
  {\bibfnamefont {S.}~\bibnamefont {Nolte}}, \bibinfo {author} {\bibfnamefont
  {K.~G.}\ \bibnamefont {Makris}}, \bibinfo {author} {\bibfnamefont
  {M.}~\bibnamefont {Segev}}, \bibinfo {author} {\bibfnamefont
  {M.}~\bibnamefont {Rechtsman}},\ and\ \bibinfo {author} {\bibfnamefont
  {A.}~\bibnamefont {Szameit}},\ }\bibfield  {title} {\bibinfo {title}
  {Topologically protected bound states in photonic parity–time-symmetric
  crystals},\ }\href {https://doi.org/10.1038/nmat4811} {\bibfield  {journal}
  {\bibinfo  {journal} {Nature Materials}\ }\textbf {\bibinfo {volume} {16}},\
  \bibinfo {pages} {433–438} (\bibinfo {year} {2017})}\BibitemShut {NoStop}%
\bibitem [{\citenamefont {Bahari}\ \emph {et~al.}(2017)\citenamefont {Bahari},
  \citenamefont {Ndao}, \citenamefont {Vallini}, \citenamefont {Amili},
  \citenamefont {Fainman},\ and\ \citenamefont {Kanté}}]{Bahari17}%
  \BibitemOpen
  \bibfield  {author} {\bibinfo {author} {\bibfnamefont {B.}~\bibnamefont
  {Bahari}}, \bibinfo {author} {\bibfnamefont {A.}~\bibnamefont {Ndao}},
  \bibinfo {author} {\bibfnamefont {F.}~\bibnamefont {Vallini}}, \bibinfo
  {author} {\bibfnamefont {A.~E.}\ \bibnamefont {Amili}}, \bibinfo {author}
  {\bibfnamefont {Y.}~\bibnamefont {Fainman}},\ and\ \bibinfo {author}
  {\bibfnamefont {B.}~\bibnamefont {Kanté}},\ }\bibfield  {title} {\bibinfo
  {title} {Nonreciprocal lasing in topological cavities of arbitrary
  geometries},\ }\href {https://doi.org/10.1126/science.aao4551} {\bibfield
  {journal} {\bibinfo  {journal} {Science}\ }\textbf {\bibinfo {volume}
  {358}},\ \bibinfo {pages} {636} (\bibinfo {year} {2017})},\ \Eprint
  {https://arxiv.org/abs/https://www.science.org/doi/pdf/10.1126/science.aao4551}
  {https://www.science.org/doi/pdf/10.1126/science.aao4551} \BibitemShut
  {NoStop}%
\bibitem [{\citenamefont {St-Jean}\ \emph {et~al.}(2017)\citenamefont
  {St-Jean}, \citenamefont {Goblot}, \citenamefont {Galopin}, \citenamefont
  {Lemaître}, \citenamefont {Ozawa}, \citenamefont {Le~Gratiet}, \citenamefont
  {Sagnes},\ and\ \citenamefont {Bloch}}]{St-Jean17}%
  \BibitemOpen
  \bibfield  {author} {\bibinfo {author} {\bibfnamefont {P.}~\bibnamefont
  {St-Jean}}, \bibinfo {author} {\bibfnamefont {V.}~\bibnamefont {Goblot}},
  \bibinfo {author} {\bibfnamefont {E.}~\bibnamefont {Galopin}}, \bibinfo
  {author} {\bibfnamefont {A.}~\bibnamefont {Lemaître}}, \bibinfo {author}
  {\bibfnamefont {T.}~\bibnamefont {Ozawa}}, \bibinfo {author} {\bibfnamefont
  {L.}~\bibnamefont {Le~Gratiet}}, \bibinfo {author} {\bibfnamefont
  {I.}~\bibnamefont {Sagnes}},\ and\ \bibinfo {author} {\bibfnamefont
  {J.}~\bibnamefont {Bloch}},\ }\bibfield  {title} {\bibinfo {title} {Lasing in
  topological edge states of a one-dimensional lattice},\ }\href
  {https://doi.org/10.1038/s41566-017-0006-2} {\bibfield  {journal} {\bibinfo
  {journal} {Nature Photonics}\ }\textbf {\bibinfo {volume} {11}},\ \bibinfo
  {pages} {651} (\bibinfo {year} {2017})}\BibitemShut {NoStop}%
\bibitem [{\citenamefont {Lieu}(2018{\natexlab{a}})}]{Lieu18a}%
  \BibitemOpen
  \bibfield  {author} {\bibinfo {author} {\bibfnamefont {S.}~\bibnamefont
  {Lieu}},\ }\bibfield  {title} {\bibinfo {title} {Topological phases in the
  non-hermitian su-schrieffer-heeger model},\ }\href
  {https://doi.org/10.1103/PhysRevB.97.045106} {\bibfield  {journal} {\bibinfo
  {journal} {Phys. Rev. B}\ }\textbf {\bibinfo {volume} {97}},\ \bibinfo
  {pages} {045106} (\bibinfo {year} {2018}{\natexlab{a}})}\BibitemShut
  {NoStop}%
\bibitem [{\citenamefont {Parto}\ \emph {et~al.}(2018)\citenamefont {Parto},
  \citenamefont {Wittek}, \citenamefont {Hodaei}, \citenamefont {Harari},
  \citenamefont {Bandres}, \citenamefont {Ren}, \citenamefont {Rechtsman},
  \citenamefont {Segev}, \citenamefont {Christodoulides},\ and\ \citenamefont
  {Khajavikhan}}]{Parto18}%
  \BibitemOpen
  \bibfield  {author} {\bibinfo {author} {\bibfnamefont {M.}~\bibnamefont
  {Parto}}, \bibinfo {author} {\bibfnamefont {S.}~\bibnamefont {Wittek}},
  \bibinfo {author} {\bibfnamefont {H.}~\bibnamefont {Hodaei}}, \bibinfo
  {author} {\bibfnamefont {G.}~\bibnamefont {Harari}}, \bibinfo {author}
  {\bibfnamefont {M.~A.}\ \bibnamefont {Bandres}}, \bibinfo {author}
  {\bibfnamefont {J.}~\bibnamefont {Ren}}, \bibinfo {author} {\bibfnamefont
  {M.~C.}\ \bibnamefont {Rechtsman}}, \bibinfo {author} {\bibfnamefont
  {M.}~\bibnamefont {Segev}}, \bibinfo {author} {\bibfnamefont {D.~N.}\
  \bibnamefont {Christodoulides}},\ and\ \bibinfo {author} {\bibfnamefont
  {M.}~\bibnamefont {Khajavikhan}},\ }\bibfield  {title} {\bibinfo {title}
  {Edge-mode lasing in 1d topological active arrays},\ }\href
  {https://doi.org/10.1103/PhysRevLett.120.113901} {\bibfield  {journal}
  {\bibinfo  {journal} {Phys. Rev. Lett.}\ }\textbf {\bibinfo {volume} {120}},\
  \bibinfo {pages} {113901} (\bibinfo {year} {2018})}\BibitemShut {NoStop}%
\bibitem [{\citenamefont {Harari}\ \emph {et~al.}(2018)\citenamefont {Harari},
  \citenamefont {Bandres}, \citenamefont {Lumer}, \citenamefont {Rechtsman},
  \citenamefont {Chong}, \citenamefont {Khajavikhan}, \citenamefont
  {Christodoulides},\ and\ \citenamefont {Segev}}]{Harari18}%
  \BibitemOpen
  \bibfield  {author} {\bibinfo {author} {\bibfnamefont {G.}~\bibnamefont
  {Harari}}, \bibinfo {author} {\bibfnamefont {M.~A.}\ \bibnamefont {Bandres}},
  \bibinfo {author} {\bibfnamefont {Y.}~\bibnamefont {Lumer}}, \bibinfo
  {author} {\bibfnamefont {M.~C.}\ \bibnamefont {Rechtsman}}, \bibinfo {author}
  {\bibfnamefont {Y.~D.}\ \bibnamefont {Chong}}, \bibinfo {author}
  {\bibfnamefont {M.}~\bibnamefont {Khajavikhan}}, \bibinfo {author}
  {\bibfnamefont {D.~N.}\ \bibnamefont {Christodoulides}},\ and\ \bibinfo
  {author} {\bibfnamefont {M.}~\bibnamefont {Segev}},\ }\bibfield  {title}
  {\bibinfo {title} {Topological insulator laser: Theory},\ }\href
  {https://doi.org/10.1126/science.aar4003} {\bibfield  {journal} {\bibinfo
  {journal} {Science}\ }\textbf {\bibinfo {volume} {359}},\ \bibinfo {pages}
  {eaar4003} (\bibinfo {year} {2018})},\ \Eprint
  {https://arxiv.org/abs/https://www.science.org/doi/pdf/10.1126/science.aar4003}
  {https://www.science.org/doi/pdf/10.1126/science.aar4003} \BibitemShut
  {NoStop}%
\bibitem [{\citenamefont {Yin}\ \emph {et~al.}(2018)\citenamefont {Yin},
  \citenamefont {Jiang}, \citenamefont {Li}, \citenamefont {L\"u},\ and\
  \citenamefont {Chen}}]{Yin18}%
  \BibitemOpen
  \bibfield  {author} {\bibinfo {author} {\bibfnamefont {C.}~\bibnamefont
  {Yin}}, \bibinfo {author} {\bibfnamefont {H.}~\bibnamefont {Jiang}}, \bibinfo
  {author} {\bibfnamefont {L.}~\bibnamefont {Li}}, \bibinfo {author}
  {\bibfnamefont {R.}~\bibnamefont {L\"u}},\ and\ \bibinfo {author}
  {\bibfnamefont {S.}~\bibnamefont {Chen}},\ }\bibfield  {title} {\bibinfo
  {title} {Geometrical meaning of winding number and its characterization of
  topological phases in one-dimensional chiral non-hermitian systems},\ }\href
  {https://doi.org/10.1103/PhysRevA.97.052115} {\bibfield  {journal} {\bibinfo
  {journal} {Phys. Rev. A}\ }\textbf {\bibinfo {volume} {97}},\ \bibinfo
  {pages} {052115} (\bibinfo {year} {2018})}\BibitemShut {NoStop}%
\bibitem [{\citenamefont {Lieu}(2018{\natexlab{b}})}]{Lieu18b}%
  \BibitemOpen
  \bibfield  {author} {\bibinfo {author} {\bibfnamefont {S.}~\bibnamefont
  {Lieu}},\ }\bibfield  {title} {\bibinfo {title} {Topological symmetry classes
  for non-hermitian models and connections to the bosonic bogoliubov--de gennes
  equation},\ }\href {https://doi.org/10.1103/PhysRevB.98.115135} {\bibfield
  {journal} {\bibinfo  {journal} {Phys. Rev. B}\ }\textbf {\bibinfo {volume}
  {98}},\ \bibinfo {pages} {115135} (\bibinfo {year}
  {2018}{\natexlab{b}})}\BibitemShut {NoStop}%
\bibitem [{\citenamefont {Shen}\ \emph {et~al.}(2018)\citenamefont {Shen},
  \citenamefont {Zhen},\ and\ \citenamefont {Fu}}]{Shen18}%
  \BibitemOpen
  \bibfield  {author} {\bibinfo {author} {\bibfnamefont {H.}~\bibnamefont
  {Shen}}, \bibinfo {author} {\bibfnamefont {B.}~\bibnamefont {Zhen}},\ and\
  \bibinfo {author} {\bibfnamefont {L.}~\bibnamefont {Fu}},\ }\bibfield
  {title} {\bibinfo {title} {Topological band theory for non-hermitian
  hamiltonians},\ }\href {https://doi.org/10.1103/PhysRevLett.120.146402}
  {\bibfield  {journal} {\bibinfo  {journal} {Phys. Rev. Lett.}\ }\textbf
  {\bibinfo {volume} {120}},\ \bibinfo {pages} {146402} (\bibinfo {year}
  {2018})}\BibitemShut {NoStop}%
\bibitem [{\citenamefont {Philip}\ \emph {et~al.}(2018)\citenamefont {Philip},
  \citenamefont {Hirsbrunner},\ and\ \citenamefont {Gilbert}}]{Philip18}%
  \BibitemOpen
  \bibfield  {author} {\bibinfo {author} {\bibfnamefont {T.~M.}\ \bibnamefont
  {Philip}}, \bibinfo {author} {\bibfnamefont {M.~R.}\ \bibnamefont
  {Hirsbrunner}},\ and\ \bibinfo {author} {\bibfnamefont {M.~J.}\ \bibnamefont
  {Gilbert}},\ }\bibfield  {title} {\bibinfo {title} {Loss of hall conductivity
  quantization in a non-hermitian quantum anomalous hall insulator},\ }\href
  {https://doi.org/10.1103/PhysRevB.98.155430} {\bibfield  {journal} {\bibinfo
  {journal} {Phys. Rev. B}\ }\textbf {\bibinfo {volume} {98}},\ \bibinfo
  {pages} {155430} (\bibinfo {year} {2018})}\BibitemShut {NoStop}%
\bibitem [{\citenamefont {Chen}\ and\ \citenamefont {Zhai}(2018)}]{Chen18}%
  \BibitemOpen
  \bibfield  {author} {\bibinfo {author} {\bibfnamefont {Y.}~\bibnamefont
  {Chen}}\ and\ \bibinfo {author} {\bibfnamefont {H.}~\bibnamefont {Zhai}},\
  }\bibfield  {title} {\bibinfo {title} {Hall conductance of a non-hermitian
  chern insulator},\ }\href {https://doi.org/10.1103/PhysRevB.98.245130}
  {\bibfield  {journal} {\bibinfo  {journal} {Phys. Rev. B}\ }\textbf {\bibinfo
  {volume} {98}},\ \bibinfo {pages} {245130} (\bibinfo {year}
  {2018})}\BibitemShut {NoStop}%
\bibitem [{\citenamefont {Lee}(2016)}]{Lee16}%
  \BibitemOpen
  \bibfield  {author} {\bibinfo {author} {\bibfnamefont {T.~E.}\ \bibnamefont
  {Lee}},\ }\bibfield  {title} {\bibinfo {title} {Anomalous edge state in a
  non-hermitian lattice},\ }\href
  {https://doi.org/10.1103/PhysRevLett.116.133903} {\bibfield  {journal}
  {\bibinfo  {journal} {Phys. Rev. Lett.}\ }\textbf {\bibinfo {volume} {116}},\
  \bibinfo {pages} {133903} (\bibinfo {year} {2016})}\BibitemShut {NoStop}%
\bibitem [{\citenamefont {Martinez~Alvarez}\ \emph {et~al.}(2018)\citenamefont
  {Martinez~Alvarez}, \citenamefont {Barrios~Vargas},\ and\ \citenamefont
  {Foa~Torres}}]{Alvarez18}%
  \BibitemOpen
  \bibfield  {author} {\bibinfo {author} {\bibfnamefont {V.~M.}\ \bibnamefont
  {Martinez~Alvarez}}, \bibinfo {author} {\bibfnamefont {J.~E.}\ \bibnamefont
  {Barrios~Vargas}},\ and\ \bibinfo {author} {\bibfnamefont {L.~E.~F.}\
  \bibnamefont {Foa~Torres}},\ }\bibfield  {title} {\bibinfo {title}
  {Non-hermitian robust edge states in one dimension: Anomalous localization
  and eigenspace condensation at exceptional points},\ }\href
  {https://doi.org/10.1103/PhysRevB.97.121401} {\bibfield  {journal} {\bibinfo
  {journal} {Phys. Rev. B}\ }\textbf {\bibinfo {volume} {97}},\ \bibinfo
  {pages} {121401} (\bibinfo {year} {2018})}\BibitemShut {NoStop}%
\bibitem [{\citenamefont {Kunst}\ \emph {et~al.}(2018)\citenamefont {Kunst},
  \citenamefont {Edvardsson}, \citenamefont {Budich},\ and\ \citenamefont
  {Bergholtz}}]{Kunst18}%
  \BibitemOpen
  \bibfield  {author} {\bibinfo {author} {\bibfnamefont {F.~K.}\ \bibnamefont
  {Kunst}}, \bibinfo {author} {\bibfnamefont {E.}~\bibnamefont {Edvardsson}},
  \bibinfo {author} {\bibfnamefont {J.~C.}\ \bibnamefont {Budich}},\ and\
  \bibinfo {author} {\bibfnamefont {E.~J.}\ \bibnamefont {Bergholtz}},\
  }\bibfield  {title} {\bibinfo {title} {Biorthogonal bulk-boundary
  correspondence in non-hermitian systems},\ }\href
  {https://doi.org/10.1103/PhysRevLett.121.026808} {\bibfield  {journal}
  {\bibinfo  {journal} {Phys. Rev. Lett.}\ }\textbf {\bibinfo {volume} {121}},\
  \bibinfo {pages} {026808} (\bibinfo {year} {2018})}\BibitemShut {NoStop}%
\bibitem [{\citenamefont {Kawabata}\ \emph {et~al.}(2018)\citenamefont
  {Kawabata}, \citenamefont {Shiozaki},\ and\ \citenamefont
  {Ueda}}]{Kawabata18}%
  \BibitemOpen
  \bibfield  {author} {\bibinfo {author} {\bibfnamefont {K.}~\bibnamefont
  {Kawabata}}, \bibinfo {author} {\bibfnamefont {K.}~\bibnamefont {Shiozaki}},\
  and\ \bibinfo {author} {\bibfnamefont {M.}~\bibnamefont {Ueda}},\ }\bibfield
  {title} {\bibinfo {title} {Anomalous helical edge states in a non-hermitian
  chern insulator},\ }\href {https://doi.org/10.1103/PhysRevB.98.165148}
  {\bibfield  {journal} {\bibinfo  {journal} {Phys. Rev. B}\ }\textbf {\bibinfo
  {volume} {98}},\ \bibinfo {pages} {165148} (\bibinfo {year}
  {2018})}\BibitemShut {NoStop}%
\bibitem [{\citenamefont {Liu}\ \emph {et~al.}(2019)\citenamefont {Liu},
  \citenamefont {Zhang}, \citenamefont {Ai}, \citenamefont {Gong},
  \citenamefont {Kawabata}, \citenamefont {Ueda},\ and\ \citenamefont
  {Nori}}]{Tao19}%
  \BibitemOpen
  \bibfield  {author} {\bibinfo {author} {\bibfnamefont {T.}~\bibnamefont
  {Liu}}, \bibinfo {author} {\bibfnamefont {Y.-R.}\ \bibnamefont {Zhang}},
  \bibinfo {author} {\bibfnamefont {Q.}~\bibnamefont {Ai}}, \bibinfo {author}
  {\bibfnamefont {Z.}~\bibnamefont {Gong}}, \bibinfo {author} {\bibfnamefont
  {K.}~\bibnamefont {Kawabata}}, \bibinfo {author} {\bibfnamefont
  {M.}~\bibnamefont {Ueda}},\ and\ \bibinfo {author} {\bibfnamefont
  {F.}~\bibnamefont {Nori}},\ }\bibfield  {title} {\bibinfo {title}
  {Second-order topological phases in non-hermitian systems},\ }\href
  {https://doi.org/10.1103/PhysRevLett.122.076801} {\bibfield  {journal}
  {\bibinfo  {journal} {Phys. Rev. Lett.}\ }\textbf {\bibinfo {volume} {122}},\
  \bibinfo {pages} {076801} (\bibinfo {year} {2019})}\BibitemShut {NoStop}%
\bibitem [{\citenamefont {Yoshida}\ \emph {et~al.}(2019)\citenamefont
  {Yoshida}, \citenamefont {Kudo},\ and\ \citenamefont {Hatsugai}}]{Yoshida19}%
  \BibitemOpen
  \bibfield  {author} {\bibinfo {author} {\bibfnamefont {T.}~\bibnamefont
  {Yoshida}}, \bibinfo {author} {\bibfnamefont {K.}~\bibnamefont {Kudo}},\ and\
  \bibinfo {author} {\bibfnamefont {Y.}~\bibnamefont {Hatsugai}},\ }\bibfield
  {title} {\bibinfo {title} {Non-hermitian fractional quantum hall states},\
  }\href {https://doi.org/10.1038/s41598-019-53253-8} {\bibfield  {journal}
  {\bibinfo  {journal} {Scientific Reports}\ }\textbf {\bibinfo {volume} {9}}
  (\bibinfo {year} {2019})}\BibitemShut {NoStop}%
\bibitem [{\citenamefont {Nakagawa}\ \emph {et~al.}(2020)\citenamefont
  {Nakagawa}, \citenamefont {Tsuji}, \citenamefont {Kawakami},\ and\
  \citenamefont {Ueda}}]{Nakagawa20}%
  \BibitemOpen
  \bibfield  {author} {\bibinfo {author} {\bibfnamefont {M.}~\bibnamefont
  {Nakagawa}}, \bibinfo {author} {\bibfnamefont {N.}~\bibnamefont {Tsuji}},
  \bibinfo {author} {\bibfnamefont {N.}~\bibnamefont {Kawakami}},\ and\
  \bibinfo {author} {\bibfnamefont {M.}~\bibnamefont {Ueda}},\ }\bibfield
  {title} {\bibinfo {title} {Dynamical sign reversal of magnetic correlations
  in dissipative hubbard models},\ }\href
  {https://doi.org/10.1103/PhysRevLett.124.147203} {\bibfield  {journal}
  {\bibinfo  {journal} {Phys. Rev. Lett.}\ }\textbf {\bibinfo {volume} {124}},\
  \bibinfo {pages} {147203} (\bibinfo {year} {2020})}\BibitemShut {NoStop}%
\bibitem [{\citenamefont {Kawabata}\ \emph {et~al.}(2022)\citenamefont
  {Kawabata}, \citenamefont {Shiozaki},\ and\ \citenamefont
  {Ryu}}]{Kawabata22}%
  \BibitemOpen
  \bibfield  {author} {\bibinfo {author} {\bibfnamefont {K.}~\bibnamefont
  {Kawabata}}, \bibinfo {author} {\bibfnamefont {K.}~\bibnamefont {Shiozaki}},\
  and\ \bibinfo {author} {\bibfnamefont {S.}~\bibnamefont {Ryu}},\ }\href@noop
  {} {\bibinfo {title} {Many-body topology of non-hermitian systems}} (\bibinfo
  {year} {2022}),\ \Eprint {https://arxiv.org/abs/2202.02548} {arXiv:2202.02548
  [cond-mat.str-el]} \BibitemShut {NoStop}%
\bibitem [{\citenamefont {McClarty}\ and\ \citenamefont
  {Rau}(2019)}]{McClarty19}%
  \BibitemOpen
  \bibfield  {author} {\bibinfo {author} {\bibfnamefont {P.~A.}\ \bibnamefont
  {McClarty}}\ and\ \bibinfo {author} {\bibfnamefont {J.~G.}\ \bibnamefont
  {Rau}},\ }\bibfield  {title} {\bibinfo {title} {Non-hermitian topology of
  spontaneous magnon decay},\ }\href
  {https://doi.org/10.1103/PhysRevB.100.100405} {\bibfield  {journal} {\bibinfo
   {journal} {Phys. Rev. B}\ }\textbf {\bibinfo {volume} {100}},\ \bibinfo
  {pages} {100405} (\bibinfo {year} {2019})}\BibitemShut {NoStop}%
\bibitem [{\citenamefont {Mook}\ \emph {et~al.}(2021)\citenamefont {Mook},
  \citenamefont {Plekhanov}, \citenamefont {Klinovaja},\ and\ \citenamefont
  {Loss}}]{Mook21}%
  \BibitemOpen
  \bibfield  {author} {\bibinfo {author} {\bibfnamefont {A.}~\bibnamefont
  {Mook}}, \bibinfo {author} {\bibfnamefont {K.}~\bibnamefont {Plekhanov}},
  \bibinfo {author} {\bibfnamefont {J.}~\bibnamefont {Klinovaja}},\ and\
  \bibinfo {author} {\bibfnamefont {D.}~\bibnamefont {Loss}},\ }\bibfield
  {title} {\bibinfo {title} {Interaction-stabilized topological magnon
  insulator in ferromagnets},\ }\href
  {https://doi.org/10.1103/PhysRevX.11.021061} {\bibfield  {journal} {\bibinfo
  {journal} {Phys. Rev. X}\ }\textbf {\bibinfo {volume} {11}},\ \bibinfo
  {pages} {021061} (\bibinfo {year} {2021})}\BibitemShut {NoStop}%
\bibitem [{\citenamefont {Stepanenko}\ and\ \citenamefont
  {Gorlach}(2020)}]{Stepanenko20}%
  \BibitemOpen
  \bibfield  {author} {\bibinfo {author} {\bibfnamefont {A.~A.}\ \bibnamefont
  {Stepanenko}}\ and\ \bibinfo {author} {\bibfnamefont {M.~A.}\ \bibnamefont
  {Gorlach}},\ }\bibfield  {title} {\bibinfo {title} {Interaction-induced
  topological states of photon pairs},\ }\href
  {https://doi.org/10.1103/PhysRevA.102.013510} {\bibfield  {journal} {\bibinfo
   {journal} {Phys. Rev. A}\ }\textbf {\bibinfo {volume} {102}},\ \bibinfo
  {pages} {013510} (\bibinfo {year} {2020})}\BibitemShut {NoStop}%
\bibitem [{\citenamefont {Olekhno}\ \emph {et~al.}(2020)\citenamefont
  {Olekhno}, \citenamefont {Kretov}, \citenamefont {Stepanenko}, \citenamefont
  {Ivanova}, \citenamefont {Yaroshenko}, \citenamefont {Puhtina}, \citenamefont
  {Filonov}, \citenamefont {Cappello}, \citenamefont {Matekovits},\ and\
  \citenamefont {Gorlach}}]{Olekhno20}%
  \BibitemOpen
  \bibfield  {author} {\bibinfo {author} {\bibfnamefont {N.~A.}\ \bibnamefont
  {Olekhno}}, \bibinfo {author} {\bibfnamefont {E.~I.}\ \bibnamefont {Kretov}},
  \bibinfo {author} {\bibfnamefont {A.~A.}\ \bibnamefont {Stepanenko}},
  \bibinfo {author} {\bibfnamefont {P.~A.}\ \bibnamefont {Ivanova}}, \bibinfo
  {author} {\bibfnamefont {V.~V.}\ \bibnamefont {Yaroshenko}}, \bibinfo
  {author} {\bibfnamefont {E.~M.}\ \bibnamefont {Puhtina}}, \bibinfo {author}
  {\bibfnamefont {D.~S.}\ \bibnamefont {Filonov}}, \bibinfo {author}
  {\bibfnamefont {B.}~\bibnamefont {Cappello}}, \bibinfo {author}
  {\bibfnamefont {L.}~\bibnamefont {Matekovits}},\ and\ \bibinfo {author}
  {\bibfnamefont {M.~A.}\ \bibnamefont {Gorlach}},\ }\bibfield  {title}
  {\bibinfo {title} {Topological edge states of interacting photon pairs
  emulated in a topolectrical circuit},\ }\href
  {https://doi.org/10.1038/s41467-020-14994-7} {\bibfield  {journal} {\bibinfo
  {journal} {Nat. Comm.}\ }\textbf {\bibinfo {volume} {11}},\ \bibinfo {pages}
  {1436} (\bibinfo {year} {2020})}\BibitemShut {NoStop}%
\bibitem [{\citenamefont {Salerno}\ \emph {et~al.}(2020)\citenamefont
  {Salerno}, \citenamefont {Palumbo}, \citenamefont {Goldman},\ and\
  \citenamefont {Di~Liberto}}]{Salerno2020}%
  \BibitemOpen
  \bibfield  {author} {\bibinfo {author} {\bibfnamefont {G.}~\bibnamefont
  {Salerno}}, \bibinfo {author} {\bibfnamefont {G.}~\bibnamefont {Palumbo}},
  \bibinfo {author} {\bibfnamefont {N.}~\bibnamefont {Goldman}},\ and\ \bibinfo
  {author} {\bibfnamefont {M.}~\bibnamefont {Di~Liberto}},\ }\bibfield  {title}
  {\bibinfo {title} {Interaction-induced lattices for bound states: Designing
  flat bands, quantized pumps, and higher-order topological insulators for
  doublons},\ }\href {https://doi.org/10.1103/PhysRevResearch.2.013348}
  {\bibfield  {journal} {\bibinfo  {journal} {Phys. Rev. Research}\ }\textbf
  {\bibinfo {volume} {2}},\ \bibinfo {pages} {013348} (\bibinfo {year}
  {2020})}\BibitemShut {NoStop}%
\bibitem [{\citenamefont {Hatano}\ and\ \citenamefont
  {Nelson}(1996)}]{Hatano96}%
  \BibitemOpen
  \bibfield  {author} {\bibinfo {author} {\bibfnamefont {N.}~\bibnamefont
  {Hatano}}\ and\ \bibinfo {author} {\bibfnamefont {D.~R.}\ \bibnamefont
  {Nelson}},\ }\bibfield  {title} {\bibinfo {title} {Localization transitions
  in non-hermitian quantum mechanics},\ }\href
  {https://doi.org/10.1103/PhysRevLett.77.570} {\bibfield  {journal} {\bibinfo
  {journal} {Phys. Rev. Lett.}\ }\textbf {\bibinfo {volume} {77}},\ \bibinfo
  {pages} {570} (\bibinfo {year} {1996})}\BibitemShut {NoStop}%
\bibitem [{\citenamefont {Hatano}\ and\ \citenamefont
  {Nelson}(1998)}]{Hatano98}%
  \BibitemOpen
  \bibfield  {author} {\bibinfo {author} {\bibfnamefont {N.}~\bibnamefont
  {Hatano}}\ and\ \bibinfo {author} {\bibfnamefont {D.~R.}\ \bibnamefont
  {Nelson}},\ }\bibfield  {title} {\bibinfo {title} {Non-hermitian
  delocalization and eigenfunctions},\ }\href
  {https://doi.org/10.1103/PhysRevB.58.8384} {\bibfield  {journal} {\bibinfo
  {journal} {Phys. Rev. B}\ }\textbf {\bibinfo {volume} {58}},\ \bibinfo
  {pages} {8384} (\bibinfo {year} {1998})}\BibitemShut {NoStop}%
\bibitem [{\citenamefont {Sch{\"a}fer}\ \emph {et~al.}(2020)\citenamefont
  {Sch{\"a}fer}, \citenamefont {Fukuhara}, \citenamefont {Sugawa},
  \citenamefont {Takasu},\ and\ \citenamefont {Takahashi}}]{Schafer2020}%
  \BibitemOpen
  \bibfield  {author} {\bibinfo {author} {\bibfnamefont {F.}~\bibnamefont
  {Sch{\"a}fer}}, \bibinfo {author} {\bibfnamefont {T.}~\bibnamefont
  {Fukuhara}}, \bibinfo {author} {\bibfnamefont {S.}~\bibnamefont {Sugawa}},
  \bibinfo {author} {\bibfnamefont {Y.}~\bibnamefont {Takasu}},\ and\ \bibinfo
  {author} {\bibfnamefont {Y.}~\bibnamefont {Takahashi}},\ }\bibfield  {title}
  {\bibinfo {title} {Tools for quantum simulation with ultracold atoms in
  optical lattices},\ }\href {https://doi.org/10.1038/s42254-020-0195-3}
  {\bibfield  {journal} {\bibinfo  {journal} {Nature Reviews Physics}\ }\textbf
  {\bibinfo {volume} {2}},\ \bibinfo {pages} {411} (\bibinfo {year}
  {2020})}\BibitemShut {NoStop}%
\bibitem [{\citenamefont {Gong}\ \emph {et~al.}(2018)\citenamefont {Gong},
  \citenamefont {Ashida}, \citenamefont {Kawabata}, \citenamefont {Takasan},
  \citenamefont {Higashikawa},\ and\ \citenamefont {Ueda}}]{Zong18}%
  \BibitemOpen
  \bibfield  {author} {\bibinfo {author} {\bibfnamefont {Z.}~\bibnamefont
  {Gong}}, \bibinfo {author} {\bibfnamefont {Y.}~\bibnamefont {Ashida}},
  \bibinfo {author} {\bibfnamefont {K.}~\bibnamefont {Kawabata}}, \bibinfo
  {author} {\bibfnamefont {K.}~\bibnamefont {Takasan}}, \bibinfo {author}
  {\bibfnamefont {S.}~\bibnamefont {Higashikawa}},\ and\ \bibinfo {author}
  {\bibfnamefont {M.}~\bibnamefont {Ueda}},\ }\bibfield  {title} {\bibinfo
  {title} {Topological phases of non-hermitian systems},\ }\href
  {https://doi.org/10.1103/PhysRevX.8.031079} {\bibfield  {journal} {\bibinfo
  {journal} {Phys. Rev. X}\ }\textbf {\bibinfo {volume} {8}},\ \bibinfo {pages}
  {031079} (\bibinfo {year} {2018})}\BibitemShut {NoStop}%
\bibitem [{\citenamefont {Okuma}\ and\ \citenamefont {Sato}(2020)}]{Okuma20b}%
  \BibitemOpen
  \bibfield  {author} {\bibinfo {author} {\bibfnamefont {N.}~\bibnamefont
  {Okuma}}\ and\ \bibinfo {author} {\bibfnamefont {M.}~\bibnamefont {Sato}},\
  }\bibfield  {title} {\bibinfo {title} {Hermitian zero modes protected by
  nonnormality: Application of pseudospectra},\ }\href
  {https://doi.org/10.1103/PhysRevB.102.014203} {\bibfield  {journal} {\bibinfo
   {journal} {Phys. Rev. B}\ }\textbf {\bibinfo {volume} {102}},\ \bibinfo
  {pages} {014203} (\bibinfo {year} {2020})}\BibitemShut {NoStop}%
\bibitem [{\citenamefont {Keilmann}\ \emph {et~al.}(2011)\citenamefont
  {Keilmann}, \citenamefont {Lanzmich}, \citenamefont {McCulloch},\ and\
  \citenamefont {Roncaglia}}]{Keilmann11}%
  \BibitemOpen
  \bibfield  {author} {\bibinfo {author} {\bibfnamefont {T.}~\bibnamefont
  {Keilmann}}, \bibinfo {author} {\bibfnamefont {S.}~\bibnamefont {Lanzmich}},
  \bibinfo {author} {\bibfnamefont {I.}~\bibnamefont {McCulloch}},\ and\
  \bibinfo {author} {\bibfnamefont {M.}~\bibnamefont {Roncaglia}},\ }\bibfield
  {title} {\bibinfo {title} {Statistically induced phase transitions and anyons
  in 1d optical lattices},\ }\href {https://doi.org/10.1038/ncomms1353}
  {\bibfield  {journal} {\bibinfo  {journal} {Nature Communications}\ }\textbf
  {\bibinfo {volume} {2}},\ \bibinfo {pages} {361} (\bibinfo {year}
  {2011})}\BibitemShut {NoStop}%
\bibitem [{\citenamefont {Rapp}\ \emph {et~al.}(2012)\citenamefont {Rapp},
  \citenamefont {Deng},\ and\ \citenamefont {Santos}}]{Rapp12}%
  \BibitemOpen
  \bibfield  {author} {\bibinfo {author} {\bibfnamefont {A.}~\bibnamefont
  {Rapp}}, \bibinfo {author} {\bibfnamefont {X.}~\bibnamefont {Deng}},\ and\
  \bibinfo {author} {\bibfnamefont {L.}~\bibnamefont {Santos}},\ }\bibfield
  {title} {\bibinfo {title} {Ultracold lattice gases with periodically
  modulated interactions},\ }\href
  {https://doi.org/10.1103/PhysRevLett.109.203005} {\bibfield  {journal}
  {\bibinfo  {journal} {Phys. Rev. Lett.}\ }\textbf {\bibinfo {volume} {109}},\
  \bibinfo {pages} {203005} (\bibinfo {year} {2012})}\BibitemShut {NoStop}%
\bibitem [{\citenamefont {Greschner}\ \emph {et~al.}(2014)\citenamefont
  {Greschner}, \citenamefont {Sun}, \citenamefont {Poletti},\ and\
  \citenamefont {Santos}}]{Greschner14}%
  \BibitemOpen
  \bibfield  {author} {\bibinfo {author} {\bibfnamefont {S.}~\bibnamefont
  {Greschner}}, \bibinfo {author} {\bibfnamefont {G.}~\bibnamefont {Sun}},
  \bibinfo {author} {\bibfnamefont {D.}~\bibnamefont {Poletti}},\ and\ \bibinfo
  {author} {\bibfnamefont {L.}~\bibnamefont {Santos}},\ }\bibfield  {title}
  {\bibinfo {title} {Density-dependent synthetic gauge fields using
  periodically modulated interactions},\ }\href
  {https://doi.org/10.1103/PhysRevLett.113.215303} {\bibfield  {journal}
  {\bibinfo  {journal} {Phys. Rev. Lett.}\ }\textbf {\bibinfo {volume} {113}},\
  \bibinfo {pages} {215303} (\bibinfo {year} {2014})}\BibitemShut {NoStop}%
\bibitem [{\citenamefont {Greschner}\ and\ \citenamefont
  {Santos}(2015)}]{Greschner15}%
  \BibitemOpen
  \bibfield  {author} {\bibinfo {author} {\bibfnamefont {S.}~\bibnamefont
  {Greschner}}\ and\ \bibinfo {author} {\bibfnamefont {L.}~\bibnamefont
  {Santos}},\ }\bibfield  {title} {\bibinfo {title} {Anyon hubbard model in
  one-dimensional optical lattices},\ }\href
  {https://doi.org/10.1103/PhysRevLett.115.053002} {\bibfield  {journal}
  {\bibinfo  {journal} {Phys. Rev. Lett.}\ }\textbf {\bibinfo {volume} {115}},\
  \bibinfo {pages} {053002} (\bibinfo {year} {2015})}\BibitemShut {NoStop}%
\bibitem [{\citenamefont {Meinert}\ \emph {et~al.}(2016)\citenamefont
  {Meinert}, \citenamefont {Mark}, \citenamefont {Lauber}, \citenamefont
  {Daley},\ and\ \citenamefont {N\"agerl}}]{Meinert16}%
  \BibitemOpen
  \bibfield  {author} {\bibinfo {author} {\bibfnamefont {F.}~\bibnamefont
  {Meinert}}, \bibinfo {author} {\bibfnamefont {M.~J.}\ \bibnamefont {Mark}},
  \bibinfo {author} {\bibfnamefont {K.}~\bibnamefont {Lauber}}, \bibinfo
  {author} {\bibfnamefont {A.~J.}\ \bibnamefont {Daley}},\ and\ \bibinfo
  {author} {\bibfnamefont {H.-C.}\ \bibnamefont {N\"agerl}},\ }\bibfield
  {title} {\bibinfo {title} {Floquet engineering of correlated tunneling in the
  bose-hubbard model with ultracold atoms},\ }\href
  {https://doi.org/10.1103/PhysRevLett.116.205301} {\bibfield  {journal}
  {\bibinfo  {journal} {Phys. Rev. Lett.}\ }\textbf {\bibinfo {volume} {116}},\
  \bibinfo {pages} {205301} (\bibinfo {year} {2016})}\BibitemShut {NoStop}%
\bibitem [{\citenamefont {Wang}\ \emph {et~al.}(2020)\citenamefont {Wang},
  \citenamefont {Hu}, \citenamefont {Eggert}, \citenamefont {Fleischhauer},
  \citenamefont {Pelster},\ and\ \citenamefont {Zhang}}]{Wang20}%
  \BibitemOpen
  \bibfield  {author} {\bibinfo {author} {\bibfnamefont {T.}~\bibnamefont
  {Wang}}, \bibinfo {author} {\bibfnamefont {S.}~\bibnamefont {Hu}}, \bibinfo
  {author} {\bibfnamefont {S.}~\bibnamefont {Eggert}}, \bibinfo {author}
  {\bibfnamefont {M.}~\bibnamefont {Fleischhauer}}, \bibinfo {author}
  {\bibfnamefont {A.}~\bibnamefont {Pelster}},\ and\ \bibinfo {author}
  {\bibfnamefont {X.-F.}\ \bibnamefont {Zhang}},\ }\bibfield  {title} {\bibinfo
  {title} {Floquet-induced superfluidity with periodically modulated
  interactions of two-species hardcore bosons in a one-dimensional optical
  lattice},\ }\href {https://doi.org/10.1103/PhysRevResearch.2.013275}
  {\bibfield  {journal} {\bibinfo  {journal} {Phys. Rev. Research}\ }\textbf
  {\bibinfo {volume} {2}},\ \bibinfo {pages} {013275} (\bibinfo {year}
  {2020})}\BibitemShut {NoStop}%
\bibitem [{\citenamefont {Weitenberg}\ and\ \citenamefont
  {Simonet}(2021)}]{Weitenberg21}%
  \BibitemOpen
  \bibfield  {author} {\bibinfo {author} {\bibfnamefont {C.}~\bibnamefont
  {Weitenberg}}\ and\ \bibinfo {author} {\bibfnamefont {J.}~\bibnamefont
  {Simonet}},\ }\bibfield  {title} {\bibinfo {title} {Tailoring quantum gases
  by floquet engineering},\ }\href {https://doi.org/10.1038/s41567-021-01316-x}
  {\bibfield  {journal} {\bibinfo  {journal} {Nature Physics}\ }\textbf
  {\bibinfo {volume} {17}},\ \bibinfo {pages} {1342–1348} (\bibinfo {year}
  {2021})}\BibitemShut {NoStop}%
\bibitem [{\citenamefont {Goldman}\ and\ \citenamefont
  {Dalibard}(2014)}]{Goldman14}%
  \BibitemOpen
  \bibfield  {author} {\bibinfo {author} {\bibfnamefont {N.}~\bibnamefont
  {Goldman}}\ and\ \bibinfo {author} {\bibfnamefont {J.}~\bibnamefont
  {Dalibard}},\ }\bibfield  {title} {\bibinfo {title} {Periodically driven
  quantum systems: Effective hamiltonians and engineered gauge fields},\ }\href
  {https://doi.org/10.1103/PhysRevX.4.031027} {\bibfield  {journal} {\bibinfo
  {journal} {Phys. Rev. X}\ }\textbf {\bibinfo {volume} {4}},\ \bibinfo {pages}
  {031027} (\bibinfo {year} {2014})}\BibitemShut {NoStop}%
\bibitem [{\citenamefont {Zhang}\ and\ \citenamefont
  {Rechtsman}(2021)}]{Zhang21}%
  \BibitemOpen
  \bibfield  {author} {\bibinfo {author} {\bibfnamefont {Z.}~\bibnamefont
  {Zhang}}\ and\ \bibinfo {author} {\bibfnamefont {M.~C.}\ \bibnamefont
  {Rechtsman}},\ }\href@noop {} {}\bibinfo {howpublished} {private
  communication} (\bibinfo {year} {2021})\BibitemShut {NoStop}%
\bibitem [{\citenamefont {Takasu}\ \emph {et~al.}(2020)\citenamefont {Takasu},
  \citenamefont {Yagami}, \citenamefont {Ashida}, \citenamefont {Hamazaki},
  \citenamefont {Kuno},\ and\ \citenamefont {Takahashi}}]{Takasu20}%
  \BibitemOpen
  \bibfield  {author} {\bibinfo {author} {\bibfnamefont {Y.}~\bibnamefont
  {Takasu}}, \bibinfo {author} {\bibfnamefont {T.}~\bibnamefont {Yagami}},
  \bibinfo {author} {\bibfnamefont {Y.}~\bibnamefont {Ashida}}, \bibinfo
  {author} {\bibfnamefont {R.}~\bibnamefont {Hamazaki}}, \bibinfo {author}
  {\bibfnamefont {Y.}~\bibnamefont {Kuno}},\ and\ \bibinfo {author}
  {\bibfnamefont {Y.}~\bibnamefont {Takahashi}},\ }\bibfield  {title} {\bibinfo
  {title} {{PT-symmetric non-Hermitian quantum many-body system using ultracold
  atoms in an optical lattice with controlled dissipation}},\ }\bibfield
  {journal} {\bibinfo  {journal} {Progress of Theoretical and Experimental
  Physics}\ }\textbf {\bibinfo {volume} {2020}},\ \href
  {https://doi.org/10.1093/ptep/ptaa094} {10.1093/ptep/ptaa094} (\bibinfo
  {year} {2020}),\ \bibinfo {note} {12A110},\ \Eprint
  {https://arxiv.org/abs/https://academic.oup.com/ptep/article-pdf/2020/12/12A110/35415105/ptaa094.pdf}
  {https://academic.oup.com/ptep/article-pdf/2020/12/12A110/35415105/ptaa094.pdf}
  \BibitemShut {NoStop}%
\bibitem [{\citenamefont {Maricq}(1982)}]{Maricq82}%
  \BibitemOpen
  \bibfield  {author} {\bibinfo {author} {\bibfnamefont {M.~M.}\ \bibnamefont
  {Maricq}},\ }\bibfield  {title} {\bibinfo {title} {Application of average
  hamiltonian theory to the nmr of solids},\ }\href
  {https://doi.org/10.1103/PhysRevB.25.6622} {\bibfield  {journal} {\bibinfo
  {journal} {Phys. Rev. B}\ }\textbf {\bibinfo {volume} {25}},\ \bibinfo
  {pages} {6622} (\bibinfo {year} {1982})}\BibitemShut {NoStop}%
\bibitem [{\citenamefont {Grozdanov}\ and\ \citenamefont
  {Rakovi\ifmmode~\acute{c}\else \'{c}\fi{}}(1988)}]{Grozdanov88}%
  \BibitemOpen
  \bibfield  {author} {\bibinfo {author} {\bibfnamefont {T.~P.}\ \bibnamefont
  {Grozdanov}}\ and\ \bibinfo {author} {\bibfnamefont {M.~J.}\ \bibnamefont
  {Rakovi\ifmmode~\acute{c}\else \'{c}\fi{}}},\ }\bibfield  {title} {\bibinfo
  {title} {Quantum system driven by rapidly varying periodic perturbation},\
  }\href {https://doi.org/10.1103/PhysRevA.38.1739} {\bibfield  {journal}
  {\bibinfo  {journal} {Phys. Rev. A}\ }\textbf {\bibinfo {volume} {38}},\
  \bibinfo {pages} {1739} (\bibinfo {year} {1988})}\BibitemShut {NoStop}%
\end{thebibliography}%

\section{Supplement}

\subsection{The Effective Doublon Model}
For the interested reader, we include a more complete derivation of our effective doublon model. We consider two bosons on a lattice of $L$ sites and construct the corresponding Hilbert space in the site basis. To denote the basis states, we introduce the notation $|n_jn_kn_l...\rangle = a_j^{\dagger^{n_j}}a_k^{\dagger^{n_k}}a_l^{\dagger^{n_l}}...|0\rangle$ where the $n_j$ is the occupancy of site $j$ with any unoccupied sites suppressed for simplicity. To construct the effective model, we apply the Hamiltonian given in Eq. 1 of the main text to three different sets of basis states: $|1_j1_k\rangle$ with $|j-k|>1$, $|2_j\rangle$, and $|1_j1_{j+1}\rangle$.

For the set of states $|1_j1_k\rangle$ with $|j-k|>1$, the density dependent terms are identically zero and the Hamiltonian is effectively non-interacting. When the Hamiltonian Eq. 1 of the main text is applied to the states $|2_j\rangle$, the density operator takes the value $n_j = 1$. The first term in Eq. 1 of the main text gives $\sqrt{2}\big[-t -i\gamma_R\big]|1_j1_{j+1}\rangle$ while the second gives $\sqrt{2}\big[-t-i\gamma_L\big]|1_{j-1}1_j\rangle$, corresponding to $J_3$ and $J_4$ in Eq. 3 of the main text, respectively. The Hamiltonian acting on the final category of states, $|1_j1_{j+1}\rangle$, results in two distinct scenarios. If the particles move further apart, e.g $|1_j1_{j+1}\rangle$ goes to $|1_j1_{j+2}\rangle$, the density-dependent term drops out and the Hamiltonian is effectively that of non-interacting particles. On the other hand if the particles move together, the first term gives $\sqrt{2}\big[-t+i\gamma_R \big]|2_{j+1}\rangle$ while the second gives $\sqrt{2}\big[-t+i\gamma_L \big]|2_{j}\rangle$, which correspond to $J_2$ and $J_1$ in Eq. 3 of the main text, respectively.

These last two sets of states form the restricted basis on which we define our effective doublon model. We consider the particles to be bound when they lie on the same or adjacent sites. As such, we consider the action of the Hamiltonian on states $|1_j1_{j+1}\rangle$ such that the particles move to the same site, i.e. we assume the doublons remain bound. The second set of states form the A sublattice ($a_{j,A}^\dagger = a_j^\dagger a_j^\dagger$) of our effective SSH model while the third set forms the B sublattice ($a_{j,B}^\dagger = a_j^\dagger a_{j+1}^\dagger$). We see the emergent sublattice symmetry as the two sets always switch into each other under the action of the Hamiltonian. The resulting doublon Hamiltonian is
\begin{equation}
    H = \sum_j J_1 a^\dagger_{j,A} a_{j,B} + J_2 a^\dagger_{j+1,A}a_{j,B} + J_3 a^\dagger_{j,B}a_{j,A} + J_4a^\dagger_{j,B}a_{j+1,A}
\end{equation}
which is equivalent to Eq. 3 of the main text when one expands the spinors $\Psi_{j,k}$.

\subsection{A Consideration on our Floquet Proposal}
Here we provide a brief discussion on two subtleties of the two frequency Floquet protocol proposed in the main text. Firstly, implementing the slow modulation must be carefully considered to obtain the model proposed in Eq. 1 of the main text. The subtlety arises in the single particle hopping. The Hatano-Nelson Hamiltonian arises from the fast modulation of the three step process described in Eq. 7 of the main text on a static Bose-Hubbard model. The resulting effective Hamiltonian describes a Hatano-Nelson model since the gradient in the loss shifts left and right hoppings of the static Hamiltonian with opposite sign. For the slow modulation to truly modulate a Hatano-Nelson Hamiltonian, one needs to also modulate the hopping in the static Hamiltonian by setting $\Delta = \Delta_0 + \Delta_t\sin(\omega t)$. In this way, we can achieve an effective stroboscopic Hamiltonian corresponding to Eq. 1 of the main text. Without taking this subtlety into account, the coupling to the non-Hermitian gauge field would have the same magnitude and we would not expect a phase with non-trivial winding number.

Additionally, to obtain accurate understanding of the dynamics, one should consider the kick operator associated with fast modulation that results in a Hatano-Nelson Hamiltonian, given as
\begin{equation}
    K(t) = -\frac{2\pi}{9\omega}(\Delta_1 \sum_j a^\dagger_{j+1}a_j + a_{j+1}a^\dagger_j)
\end{equation}
which is mathematically equivalent to a kinetic energy term. In this article we have studied the energy spectrum and topology of the system, which is not sensitive to the kick operator. A full understanding of the dynamics of the system is left to future work.

\subsection{Additional Floquet Protocols}

Here we present two additional Floquet protocols that may be of interest. The first is similar to that obtained in the above paper with an additional modulated interacting term. The second protocol is derived in the formalism of Ref. [40].

\subsection{Magnus Expansion}
The first Floquet protocol we will consider is described by the time dependent Hamiltonian
\begin{equation}
\begin{split}
    H(t) =& \Delta\sum_j \bigg[a^\dagger_{j+1}a_j + a^\dagger_ja_{j+1}\bigg] + U\sum_jn_j(n_j-1)\\
    + \sin&(\omega t)\sum_j\bigg[\Delta_Ra^\dagger_{j+1}a_j + \Delta_La^\dagger_ja_{j+1} +U_\omega n_j(n_j-1)\bigg]
\end{split}
\end{equation}
which describes a Bose-Hubbard Hamiltonian with sinusoidal modulation of the hopping and on-site density-density term. From the Magnus expansion, we obtain the effective stroboscopic Hamiltonian to first order in $1/\omega$
\begin{align}
    H_{\text{eff}} &= \sum_j a^\dagger_{j+1}\bigg[\Delta+\frac{2}{i\hbar\omega} (\Delta U_\omega-U\Delta_R)(n_j-n_{j+1})\bigg]a_j \notag \\
    &+ a^\dagger_{j}\bigg[\Delta+\frac{2}{i\hbar\omega} (\Delta U_\omega-U\Delta_L)(n_{j+1}-n_{j})\bigg]a_{j+1} \notag \\
    &+ U\sum_j a_j^\dagger a_j (a_j^\dagger a_j - 1)
\end{align}
which we map to Eq. 1 of the main text by setting $\Delta = -t$, $\gamma_L = \frac{2}{\hbar\omega}(\Delta U_\omega-U\Delta_L)$ and $\gamma_R = \frac{2}{\hbar\omega}(\Delta U_\omega-U\Delta_R)$. The effective Hamiltonian will be in the topologically non trivially phase if $U$ is imaginary, corresponding to a two-body loss, and $\Delta_R,\Delta_L \in \mathbb{C}$. In this method the resulting non-Hermiticity arises from the loss in the static Hamiltonian.

\subsection{Goldman-Dalibard}
Another potential Floquet protocol for realizing our model can be obtained through a three step sequence of square waves, similar to our proposal for realizing a Hatano-Nelson model. The Floquet protocol is described by
\begin{equation}
    H = -\Delta \sum_j (a^\dagger_{j+1}a_j + a_{j+1}a^\dagger_j) + V(t)
\end{equation}
where $\Delta$ is the hopping parameter and $V(t)$ is the modulation given by
\begin{equation}
    V(t) = 
    \begin{cases}
      \Delta_1 \sum_j a^\dagger_{j+1}a_j + a_{j+1}a^\dagger_j, & 0 \le t < T/3 \\
      \Delta_2 \sum_j a^\dagger_ja_j(a^\dagger_ja_j-1), & T/3 \le t < 2T/3 \\
      i\Delta_3 \sum_j a^\dagger_{j+1}a_j - a_{j+1}a^\dagger_j, & 2T/3 \le t < T
    \end{cases}
\end{equation}
Here $T$ is the total period of the drive, time $t$ is defined mod $T$, and the $\Delta_i$ give the strength of each modulation. The first step is a modulation of the hopping parameter of the free Bose gas. The second step is an onsite two-body interaction. The final step corresponds to a linear pulse in momentum.

The effective Hamiltonian for this modulation, obtained from the formalism of Ref. [40], is given by 
\begin{align}
    H_{\text{eff}} &= \sum_j a^\dagger_{j+1}\bigg[-\Delta + \frac{i\Delta_2\pi(\Delta_1+i\Delta_3)}{27\omega}(n_{j+1} - n_{j})\bigg]a_j \notag \\
    +& a^\dagger_{j}\bigg[-\Delta + \frac{i\Delta_2\pi(\Delta_1-i\Delta_3)}{27\omega}(n_{j} - n_{j+1})\bigg]a_{j+1}.
\end{align}
To map to Eq. 1 of the main text, we set $\Delta =t$, $\gamma_L=\frac{i\Delta_2\pi(\Delta_1+i\Delta_3)}{27\omega}$, and $\gamma_R=\frac{i\Delta_2\pi(\Delta_1-i\Delta_3)}{27\omega}$. To realize the topological phases discussed above, one option is to choose $\Im[\Delta_3]\neq 0$, which will give asymmetric coupling to the density dependent gauge field.
\end{document}